\documentclass[12pt]{article}
\usepackage{cite,psfig}

\newcommand{\newc}{\newcommand}
\newc{\gsim}{\lower.7ex\hbox{$\;\stackrel{\textstyle>}{\sim}\;$}}
\newc{\lsim}{\lower.7ex\hbox{$\;\stackrel{\textstyle<}{\sim}\;$}}
\newc{\gev}{\,{\rm GeV}}
\newc{\mev}{\,{\rm MeV}}
\newc{\ev}{\,{\rm eV}}
\newc{\kev}{\,{\rm keV}}
\newc{\tev}{\,{\rm TeV}}
\def\ln{\mathop{\rm ln}}
\def\tr{\mathop{\rm tr}}

\newc{\mz}{M_Z}
\newc{\mpl}{M_*}
\newc{\mw}{m_{\rm weak}}
\def\beq{\begin{equation}}
\def\eeq{\end{equation}}
\def\bea{\begin{eqnarray}}
\def\eea{\end{eqnarray}}
\newc{\ie}{{\it i.e.}}          \newc{\etal}{{\it et al.}}
\newc{\eg}{{\it e.g.}}          \newc{\etc}{{\it etc.}}
\newc{\cf}{{\it c.f.}}
\def\bar#1{\overline{#1}}
\def\vev#1{\left\langle #1 \right\rangle}

\def\inv{^{\raise.15ex\hbox{${\scriptscriptstyle -}$}\kern-.05em 1}}
\def\lbar{{\lower.35ex\hbox{$\mathchar'26$}\mkern-10mu\lambda}} 

\let\<=\langle
\let\>=\rangle

\let\+=\uparrow
\let\al=\alpha

\let\de=\delta

\let\la=\lambda
\let\La=\Lambda

\let\si=\sigma
\let\Si=\Sigma
\let\th=\theta

\renewcommand{\epsilon}{\varepsilon}
\renewcommand{\phi}{\varphi}

\begin{document}
\thispagestyle{empty}
\vspace*{.5cm}
\noindent
\hspace*{\fill}{CERN-TH/2001-156}\\
\hspace*{\fill}{LBNL-48220}
\vspace*{2.0cm}

\begin{center}
{\Large\bf A Minimal $S^1/(Z_2\times Z_2')$ Orbifold GUT}
\\[2.5cm]
{\large Arthur Hebecker$^*$ and John March-Russell$^{*,\dagger}$
}\\[.5cm]
{\it $^*$Theory Division, CERN, CH-1211 Geneva 23, Switzerland}
\\[.2cm]
{\it $^\dagger$Theory Group, LBNL, University of California,
Berkeley, CA 94720, USA}
\\[.2cm]
(June 15, 2001)
\\[1.1cm]

{\bf Abstract}\end{center}
\noindent
We investigate supersymmetric SU(5) grand-unified theories (GUTs)
realized in 5 space-time dimensions and broken down to the MSSM by
SU(5)-violating boundary conditions on a $S^1/(Z_2\times Z_2')$ orbifold
with two 3-branes.  The doublet-triplet splitting problem is entirely
avoided by locating the MSSM Higgs doublets on the brane on which
SU(5) is not a good symmetry.  An extremely simple model is then described
in which the MSSM matter is also located on this SU(5)-violating brane.
Although this model does not unify the MSSM matter within SU(5) multiplets,
it explains gauge coupling unification. A second model with MSSM matter
in the SU(5)-symmetric bulk preserves both the SU(5) explanation of fermion
quantum numbers as well as gauge-coupling unification.  Both models
naturally avoid problematic SU(5) predictions for the Yukawa couplings of
the first two generations and are consistent with proton decay constraints.
We analyse the running of gauge couplings above the compactification scale
in terms of a 5d effective action and derive the implications for the
values of compactification scale, unification scale and of the scale at
which the bulk gauge theory becomes strongly coupled.
\newpage

\setcounter{page}{1}
\section{Introduction}

Since the pioneering work of Georgi and Glashow~\cite{GG} (also
\cite{PS}) the compelling concept of a grand unified theory (GUT)
of the Standard Model (SM) gauge interactions has dominated our thinking about
physics at high energies. The success of gauge coupling unification~\cite{GQW}
in minimal supersymmetric standard model (MSSM) extensions of these
theories~\cite{DRW,DG}
has further supported this idea~\cite{unif}.  However, despite these
great successes (and others, such as the generation of small neutrino
masses) this GUT concept is not without its faults. In particular we
recall the problems of the Higgs structure of the high-scale theory,
especially the doublet-triplet splitting problem, the issue of too fast
proton decay, and the mismatch of the GUT scale with the naive scale of
unification with gravity.

Recently, a new possibility for the embedding of the SM into a form of
GUT has been suggested by Kawamura~\cite{kaw1,kaw2,kaw3} and further
extended by Altarelli and Feruglio~\cite{AF} and Hall and
Nomura~\cite{HN}.  The basic idea is
that the GUT gauge symmetry is realized in 5 or more space-time
dimensions and only broken down to the SM by utilizing GUT-symmetry
violating boundary conditions on a singular `orbifold' compactification.
Given the success of the traditional supersymmetric gauge-coupling
unification predictions, the most attractive class of models are ones
with both supersymmetry and (at least) SU(5) gauge symmetry in 5
dimensions.  In this case both the GUT group and 5d supersymmetry
are broken down to a N=1 supersymmetric model with
SM gauge group by compactification on $S^1/(Z_2\times Z_2')$ (a related
idea was employed for EWSB in Ref.~\cite{BHN}).
This construction allows one to avoid some unsatisfactory features
of conventional GUTs with Higgs breaking, such as doublet-triplet
splitting, while maintaining, at least at leading order, the desired
MSSM gauge coupling unification.  Moreover, given the fact that string
theory requires additional dimensions, as well as branes located at
orbifold fixed points, the necessary (inverse-GUT-scale-sized) 5th
dimension is not unreasonable.\footnote{We will take an
effective-field-theory viewpoint of orbifold branes, namely we will
impose all constraints necessary for a consistent low-energy theory,
but not require an explicit string-theory (or other quantum gravity)
realisation.}

In this letter we present two GUT models of this type, which we
believe to be the simplest yet discovered, our particular focus
being the realization of the MSSM Higgs and the MSSM matter.
Specifically in Sect.~2 we review the basic physics of the
construction of Refs.~\cite{kaw1,kaw2,kaw3,AF,HN}, applying this in
Sect.~2.1 to the gauge sector of the model. Sect.~2.2
contains the novel observation that the electroweak Higgs can be
a localized field on the SU(5)-violating brane, with no triplet
partners whatsoever, thus solving the doublet-triplet problem of
traditional GUTs in a particularly simple way.  Sect.~2.3
contains a detailed discussion of two novel constructions of the
MSSM matter in such theories.  The first of these two models
locates the MSSM matter on the SU(5)-violating
brane and has both an exceptionally simple structure and is
evidently consistent with low-energy phenomenology.  The
model explains gauge coupling unification but, intriguingly,
has no SU(5) multiplet structure for the MSSM matter, so in
itself does not explain the quantum numbers of the MSSM matter.
In our second model we present a novel way of realising the MSSM spectrum
from SU(5)-representations in the 5d bulk (thus explaining the
matter quantum numbers), which nevertheless leads to zero-modes
which couple to the brane-localized Higgs in a way that does not
imply the incorrect SU(5) relations for the Yukawa matrices
for the first two generations.  Thus both models possess
a number of appealing features.
In Sect.~3,  we address the nature of gauge coupling unification
in these models (developing the treatment of~\cite{HN}), with special
emphasis on the corrections to exact unification arising from
SU(5)-violating brane kinetic terms.  We analyse
the running of gauge couplings above the compactification scale in
terms of a 5d effective
action and derive the implications for the values of the compactification
scale, the unification scale and of the scale at which the bulk gauge theory
becomes strongly coupled.  Most of the discussion of Sect.~3 is generic
and applies to other models with orbifold GUT breaking as well.
Sect.~4 contains our conclusions.

\section{The Structure of the Minimal Model(s)}

Let us recall in greater detail the essential features of the
$S^1/(Z_2\times Z_2')$ model as formulated in Refs.~\cite{kaw2,AF,HN}.
Even at this simple stage we will argue that there are more possibilities
for the construction of the theory than have been utilized so far,
and that the new model(s) we describe more minimally solve the problems
of traditional 4d GUTs.

Consider a 5-dimensional factorized space-time comprising
a product of 4d Minkowski space M$_4$ (with coordinates $x^{\mu}$,
$\mu=0,\ldots,3$), and the orbifold $S^1/Z_2\times Z_2'$, with coordinate
$y\equiv x^5$.  The circle $S^1$ will be taken to have radius
$R$ where $1/R\sim M_{\rm GUT}$ (so large extra dimensions are not being
considered).  The orbifold $S^1/Z_2$ is obtained by modding out the theory
by a $Z_2$ transformation which imposes on fields which depend upon
the 5th coordinate the equivalence relation $y\sim -y$.
`Modding out' the theory means that we restrict the configuration
space of fields in the functional integral to ones that are invariant
under the specified $Z_2$ action.
To obtain the orbifold $S^1/(Z_2\times Z_2')$ we further mod out
by $Z'_2$ which imposes the equivalence relation $y'\sim -y'$,
with $y'\equiv y+\pi R/2$.
Actually this formulation of the $S^1/(Z_2\times Z_2')$ orbifold
slightly obscures one fact:  The second equivalence
$y'\sim -y'$ can be combined with the first, $y\sim -y$, to obtain the
physically identical equivalence $y\sim y +\pi R$, and this slight
re-writing makes manifest that the second equivalence
introduces no new fixed points by itself.  However, let us continue
to work with the basis of identifications
\beq
P:~~y\sim -y \qquad P':~~y' \sim -y' .
\label{eq:PPequiv}
\eeq
It is important to the whole construction that under these equivalences
there are two inequivalent fixed 3-branes (or `orbifold planes')
located at $y=0$, and $y =\pi R/2\equiv\ell$, which we denote $O$ and $O'$
respectively.  It is consistent to work with the theory obtained by
truncating to the physically irreducible interval
$y\in [0,\ell]$ with the 3-branes at $y=0,\ell$ acting as
`end-of-the-world' branes, and henceforth we so do.

The action of the equivalences $P,P'$ on the fields of a quantum field
theory living on $M_4\times S^1/(Z_2\times Z_2')$ is not fully
specified by the action Eq.~(\ref{eq:PPequiv}) on the coordinates.  One
must also define the action within the space of fields.  To this end, let
$\Phi(x,y)$ be a vector comprising all bulk fields, then the action
of $P$ and $P'$ is given by
\bea
P: \Phi(x,y)&\sim & P_\Phi \Phi(x,-y),\\
P': \Phi(x,y')&\sim & P'_\Phi \Phi(x,-y').
\eea
Here, on the rhs, $P_\Phi$ and $P'_\Phi$ are matrix representations
of the two $Z_2$ operator actions, which necessarily
have eigenvalues $\pm 1$ when diagonalized.
Let us from now on work in this diagonal basis of fields, and
classify the fields by their eigenvalues $(\pm 1,\pm 1)$. Then
these fields have the KK expansions:
\bea
\Phi_{++}(x,y) &=& \sqrt{4 \over \pi R}
\sum_{n=0}^\infty
{1\over(\sqrt{2})^{\de_{n,0}}} \Phi^{(2n)}_{++}(x)\cos{2ny\over R},
\label{eq:++}\\
\Phi_{+-}(x,y) &=& \sqrt{4\over \pi R}
\sum_{n=0}^\infty \Phi^{(2n+1)}_{+-}(x)\cos{(2n+1)y \over R},
\label{eq:+-}\\
\Phi_{-+}(x,y) &=& \sqrt{4 \over \pi R}
\sum_{n=0}^\infty \Phi^{(2n+1)}_{-+}(x)\sin{(2n+1)y \over R},
\label{eq:-+}\\
\Phi_{--}(x,y) &=& \sqrt{4 \over \pi R}
\sum_{n=0}^\infty \Phi^{(2n+2)}_{--}(x) \sin{(2n+2)y \over R}.
\label{eq:--}
\eea
From the 4d perspective the KK fields
$\Phi^{(k)}(x)$ acquire a mass $k/R$, so only the $\Phi_{++}$ possess
a massless zero mode.  Moreover, only $\Phi_{++}$ and $\Phi_{+-}$
have non-zero values at $y=0$, while only $\Phi_{++}$ and $\Phi_{-+}$
are non-vanishing at $y=\ell$.

It is important to realize that the action of the identifications
$P,P'$ on the fields (namely the matrices $P_\Phi$ and $P'_\Phi$)
can utilize {\em all} of the symmetries of the
bulk theory.  Thus $P$ and $P'$ can involve gauge transformations,
discrete parity transformations, and most importantly in the supersymmetric
case, R-symmetry transformations.  This last feature allows one to
break the higher-dimensional supersymmetry to a phenomenologically
acceptable, and desirable N=1 SUSY theory in 4d.

\subsection{The Gauge Structure of the Minimal Model}

To reproduce the good predictions of a minimal supersymmetric GUT,
we start from a 5d SU(5) gauge theory with minimal SUSY in 5d (with 8 real
supercharges, corresponding to N=2 SUSY in 4d).  Thus, at minimum, the
bulk must have the 5d vector superfield, which in terms of 4d N=1 SUSY
language contains a vector supermultiplet $V$ with physical components
$A_\mu,\la$, and a chiral multiplet $\Si$ with components $\psi,\si$.
Both $V$ and $\Si$ transform in the adjoint representation of SU(5).

Now choose the matrix representation of the parity assignments,
expressed in the fundamental representation of SU(5), to be
\beq
P={\rm diag}(+1,+1,+1,+1,+1), \qquad P'={\rm diag}(-1,-1,-1,+1,+1),
\label{eq:PP}
\eeq
so that
\bea
P: V^a(x,y)T^a &\sim & V^a(x,-y)~ P T^a P^{-1},\nonumber\\
P': V^a(x,y')T^a &\sim & V^a(x,-y')~ P' T^a P'^{-1} ,
\label{eq:PonV}
\eea
where the action on the rhs is as matrices.  For $\Si$ the same
assignments are taken apart from an overall sign for both $P$ and $P'$,
so that
\bea
P: \Si^a(x,y)T^a & \sim & - \Si^a(x,-y)~ P T^a P^{-1},\nonumber\\
P': \Si^a(x,y)T^a & \sim & - \Si^a(x,-y)~ P' T^a P'^{-1} .
\label{eq:PonSi}
\eea
These boundary conditions then break
SU(5) to the SM gauge group on the $O'$ brane at $y=\ell$,
and 4d N=2 SUSY to 4d N=1 SUSY on both the $O$ and $O'$ branes.
This can be seen explicitly by examining the masses of the KK towers
of the fields as displayed in Table~1.  Only the $(+,+)$ fields possess
massless zero modes, and at low energies the gauge and gaugino content
of the 4d N=1 MSSM is apparent.
{\begin{center}
\begin{tabular}{|c|c|c|}
\hline
$(P,P')$ & 4d superfield & 4d mass\\
\hline
$(+,+)$ &  $V^a$ & $2n/R$\\
$(+,-)$ &  $V^{\hat{a}}$ & $(2n+1)/R$ \\
$(-,+)$ &  $\Si^{\hat{a}}$ & $(2n+1)/R$\\
$(-,-)$ &  $\Si^{a}$ & $(2n+2)/R$ \\
\hline
\end{tabular}
\end{center}}
\vspace{3mm}
Table 1. Parity assignment and KK masses of fields in the 4d vector
and chiral adjoint supermultiplet. The index $a$ labels the
unbroken SU(3)$\times$SU(2)$\times$U(1) generators of SU(5),
while $\hat{a}$ labels the broken generators.

\subsection{The Electroweak Higgs}

We now come to the first significant difference with respect to the
treatments of Refs.~\cite{kaw1,kaw2,kaw3,AF,HN}: our model does not include
bulk Higgs multiplets.  If we were to include such bulk Higgs, then
the unbroken SU(5) gauge invariance of the {\em bulk} would force
them to transform as ${\bf 5}$'s (and $\bar {\bf 5}$'s) of SU(5).
As such representations contain more than the SU(2) doublets
necessary for electroweak symmetry breaking, we are faced with
a version of the infamous doublet-triplet
splitting problem.  Of course a success of the models of
Refs.~\cite{kaw1,kaw2,kaw3,AF,HN} is that by extending the action of $P,P'$
to this bulk Higgs sector, the triplet components can be made heavy.
Nevertheless it is unarguably simpler (and phenomenologically different)
never to have had the Higgs triplet components in the first place!

How then are the Higgs doublets to be realized?  To understand the
possibilities it is important to realize that after the orbifolding,
Eqs.~(\ref{eq:PP})-(\ref{eq:PonSi}), the amount
of unbroken gauge symmetry is {\em position dependent}.  On the $O$ brane
at $y=0$ and in the bulk the entire SU(5) symmetry is a good
symmetry.  In particular at position $y\neq\ell$ the allowed gauge
transformations are of the form $U=\exp(i \sum_{a}\xi^a(x,y) T^a + i
\sum_{\hat{a}} \xi^{\hat{a}}(x,y) T^{\hat{a}} )$, with both the gauge
transformation parameters $\xi^a(x,y)$ and $\xi^{\hat{a}}(x,y)$
non-vanishing.  Only at the $O'$ brane are the
$\xi^{\hat{a}}(x,\ell) = 0$, and purely the SM gauge symmetry is defined.
The implication of this for the gauge sector of the theory
is that the form of the 5d effective Lagrangian (at scale $\mu$) is:
\beq
{\cal L}_{\rm eff}(\mu)=\, \int\, d^2\th
\left\{-\left(\frac{1}{2g_5(\mu)^2}+\frac{\de(y)}{2g_0(\mu)^2}\right)
\tr {\cal W}^\al {\cal W}_\al
-\sum_{i=1}^3\frac{\de(y-\ell)}{2g_i(\mu)^2}
\tr{\cal W}^\al_i {\cal W}^i_\al\right\} ,
\label{eq:effL}
\eeq
where ${\cal W}_{\al,i}$ ($i=1,2,3$) are the supersymmetric
field strengths of the unbroken
U(1)$\times$SU(2)$\times$SU(3) SM gauge group on the $O'$ brane,
and ${\cal W}$ is the SU(5) field strength.  In other words, the effective
action of the $O'$ brane need only respect the SM gauge group, while
the bulk theory must respect the full SU(5) gauge symmetry.

The symmetries of the theory allow one to add interactions, and
additional matter, on the $O'$ brane which only respect the SM gauge
symmetry.  Thus, unlike the models of Refs.~\cite{kaw1,kaw2,kaw3,AF,HN},
we choose to add a pair of
weak-SU(2) doublet N=1 chiral superfields $H_u,H_d$ to the $O'$ 3-brane
theory at $y=\ell$.  This is a particularly natural
and attractive definition of the theory precisely because the Higgs
is the one matter field in the MSSM that does not have partners
that effectively fill out a full SU(5) multiplet as far as quantum
numbers are concerned.  As we do not have any Higgses which transform
as a triplet of SU(3), there is automatically no doublet-triplet-splitting
problem either.

\subsection{The MSSM Matter}

The next question is the location of the SM matter (and their
4d N=1 superpartners). In our framework it cannot be on the
SU(5)-invariant $O$ brane at $y=0$ as it would not then be able to interact
with the electroweak Higgs. This leaves us with two possibilities, both
of which lead to attractive models.

\noindent
{\bf I. SM matter on the SM brane}

If, together with the Higgs, the matter is also located on the
$O'$ brane, then the Yukawa couplings only need be
invariant under the SM gauge group.
Thus, the Yukawa couplings are no more restricted than in the traditional
4d MSSM, and automatically there is enough flexibility to accommodate fermion
masses and mixings.  Therefore the Yukawa terms in the superpotential
of the 5d effective lagrangian for the light SM-charged
fields of our minimal theory are localized entirely on the $O'$ brane
at $y=\ell$:
\beq
W_{\rm Yukawa}=\int d^2\th \de(y-\ell)\sum_{IJ}\biggl(
h^u_{IJ} Q_I H_u \bar{U}_J +h^d_{IJ} Q_I H_d \bar{D}_J
+ h^l_{IJ} L_I H_d \bar{E}_J \biggr),
\label{eq:YukI}
\eeq
where $I,J=1,2,3$ are generation indices, and all fields depend only
upon $x^\mu$.
This theory has considerable advantages as compared to the set-ups of both
Altarelli and Feruglio~\cite{AF} (AF) and Hall and Nomura~\cite{HN} (HN).
Specifically, the construction of AF considers the Higgs to be a bulk field
(which therefore must be extended to a full ${\bf 5}$), and locates the SM
on the SU(5) invariant brane (the SM matter therefore transform as full
$\bar{\bf 5} + {\bf 10}$ multiplets of SU(5)).  On this
brane AF then assume {\em SU(5)-non-invariant} Yukawa interactions
between the bulk Higgs and the SM matter.  But such a set-up is not
consistent with unbroken gauge invariance on the $O$ brane. In particular
it does not allow the freedom to move between unitary and covariant gauges
to show the simultaneous unitarity and Lorentz-invariance of the low-energy
theory.  (Lack of renormalizability, at least at high scales, is not
really the important issue -- after all the basic Kawamura set-up considers
a 5d non-renormalizable theory.  This is perfectly OK if the scale of the
loss of calculability is sufficiently high.)
Indeed unless one includes further SU(5)-breaking mass terms
for the SU(5) gauge bosons, $m^2 A_\mu^{\hat a} A^{\mu,{\hat a}}$
(localized to the formerly SU(5)-invariant brane), the theory
is not unitary at low energy.  The whole theory then has a rather different,
and much more complicated character than that discussed by Altarelli
and Feruglio.  We therefore consider this theory to be more complicated
than necessary.

On the other hand the theory of HN, where again the Higgs
is taken to be a bulk field, and the SM matter is located on the
SU(5)-invariant brane, assumes SU(5) invariant Yukawa interactions
localized to the SU(5)-invariant brane. This is a consistent set-up.
Unlike in our model where we place the SM matter on the $O'$ brane, the
HN model can explain the $b$-$\tau$ mass ratio~\cite{BEGN} via SU(5)
relations for
the Yukawas.  Unfortunately it also leads to SU(5) predictions for the
first and second generation fermion mass ratios, $m_d/m_e$ and $m_s/m_\mu$,
which are in strong disagreement with experiment.  Thus HN are forced to
introduce additional fields in the bulk which mix with the first and second 
generations of MSSM fields so as to correct the wrong predictions for the 
first two generations of quarks and leptons. Although possible, this 
procedure is {\em ad hoc}.

Our model does not suffer from either of the above shortcomings.  It is 
also physically different in advantageous ways.  First, since there are no 
Higgs triplets at all, it immediately follows that there are no problems 
with triplet-Higgsino-mediated proton decay.  (The models of AF and HN, 
where triplet Higgsinos exist as KK excitations, also avoid dangerous 
Higgsino-mediated proton decay, but for different reasons.) There are no 
dimension-5 proton decay mediating operators
from Higgsinos in our theory.  Of course, there are still proton-decay
processes mediated by the heavy $(+,-)$ KK mode $X$ and $Y$ gauge
bosons.  These are present due to terms in the $O'$-brane action
involving the (non-vanishing) $\partial_5$ derivatives of
$A_{\mu}^{\hat a}$. (Note that such KK excitations in the $X$ and $Y$
gauge boson directions carry hypercharge, which is correctly normalized
because it sits inside the bulk SU(5), so their interactions with the
brane-localized SM states also correctly normalize the hypercharge of
these SM states.)  Being dimension 6 operators these lead to much
less of a constraint on the effective GUT scale $1/R$. In addition,
however, there can in principle be brane-localized dimension 5
baryon-number-violating operators from global-quantum-number-violating
quantum gravity effects~\cite{qgravity}.  Because of the existence
of a relatively large 5th dimension the quantum gravity scale is
lower than the usual Planck mass, so these operators, if they are there,
are more problematic than usual.  We leave a detailed discussion
of this issue to a future publication.
Second, this model has several very nice features when one considers
the breaking of the remaining N=1 supersymmetry down to the SM at
the TeV scale.  An attractive solution of the supersymmetric
FCNC problem follows if the MSSM matter is sequestered from
the SUSY-breaking sector.  The localization of the MSSM matter
to the $O'$-brane naturally allows us to achieve this
separation by having the SUSY-breaking sector on the SU(5)-symmetric
$O$-brane, and communicating the SUSY breaking to the MSSM matter
by the MSSM gauginos which live in the bulk.  This is similar but not
identical to the models of~\cite{gaugino} considered by HN~\cite{HN}.  For 
example, one difference arises from the
fact that in our model the gauginos necessarily pick up their soft
SUSY-breaking mass from an SU(5)-symmetric interaction
(\eg, $(S\tr {\cal W}^\al {\cal W}_\al)|_{F-{\rm cpt}}$ if there
is a singlet chiral superfield $S$ on $O$ with $\vev{F_S}\neq 0$).
Thus the gaugino masses will satisfy the traditional SUSY SU(5)
relations to the same high degree of accuracy as the gauge-couplings
themselves unify.  This and other details of SUSY breaking in this
(and the following) model will be discussed in a future publication.
An aesthetic disadvantage of this model is that it gives up on an
explanation via SU(5) multiplet structure of the quantum numbers of
the SM matter fields. (The well-known restrictions of anomaly cancellation
for these quantum numbers of course still apply.)  Overall we find this
model intriguing because of its extreme simplicity and because it manages
to unify the gauge couplings without a corresponding unification of the
Higgs or MSSM matter.

\noindent
{\bf II. SM matter in the bulk}

We now discuss an alternate model for the MSSM matter.
The bulk of our theory is invariant under both the full SU(5) gauge
symmetry and the full `N=2' minimal supersymmetry (8 real supercharges)
of 5 dimensions.  If we take the MSSM matter to reside in the bulk, then
the first of these symmetries immediately tells us that they must come
in complete SU(5) multiplets.  Therefore the usual SU(5) matter
multiplet structure, ${\bf 10} + {\bf \bar{5}}$, is naturally
incorporated, and the quantum numbers of the SM matter fields, in particular
the hypercharge assignments and the issue of charge quantization, are
explained for the same reasons as in traditional 4d GUTs with simple
gauge group.  In fact we will soon argue that the correct situation with
regard to quantum numbers is slightly more subtle in a particularly
interesting fashion.

On the other hand, the bulk N=2 SUSY appears to pose a problem.  The minimal
matter superfield representation for such a theory is a hypermultiplet,
which in 4d N=1 language decomposes in to a chiral multiplet $\Phi_{\bf R}$
together with a mirror chiral multiplet in the conjugate representation
$\Phi^c_{\bf \bar{R}}$. Thus, the choice of matter in the bulk would
appear to have problems reproducing the chiral structure of the SM.

Fortunately we are saved once again by the structure of the orbifold
projections $P$ and $P'$ acting on fields.  The action of these
projections on the N=1 component fields $\Phi$ and $\Phi^c$ residing in a
5d hypermultiplet is inherited from the action on the 5d vector
multiplet Eqs.~(\ref{eq:PP})-(\ref{eq:PonSi}).
The result is that actions of both $P$ and $P'$ on the 4d chiral
fields $\Phi$ and $\Phi^c$ have a relative sign:
\beq
P: \Phi \sim P_\Phi\Phi , \qquad P: \Phi^c \sim -P_\Phi\Phi^c\,,
\label{eq:Ponhypers}
\eeq
and similarly for $P'$.  This difference leads to a chiral spectrum
for the zero modes.  Indeed the KK spectrum of 5d bulk hypermultiplets
in the representation ${\bf 10} + {\bf \bar{5}}$ (whose 4d chiral
components we denote $T+\bar{F}+T^c + \bar{F}^c$) resulting from
the $P,P'$ actions is given in Table~2.  Note that since $P'$
does not commute with SU(5), components of the $T+\bar{F}+T^c +
\bar{F}^c$ fields in different SU(3)$\times$SU(2)$\times$U(1)
representations have different KK mode structures.  Thus in this
table we label the expansions by $Q,\bar{U},\bar{D},L,\bar{E}$, etc.,
to indicate their SM transformation properties.

{\begin{center}
\begin{tabular}{|c|c|c|}
\hline
$(P,P')$ & 4d superfield & 4d mass\\
\hline
$(+,+)$ &  $T_{\bar U}, T_{\bar E}, \bar{F}_{L}$  & $2n/R$\\
$(+,-)$ &  $T_{Q}, \bar{F}_{\bar D} $ & $(2n+1)/R$ \\
$(-,+)$ &  $T^c_{\bar Q}, \bar{F}^c_{D}$ & $(2n+1)/R$\\
$(-,-)$ &  $T^c_{U}, T^c_{E}, \bar{F}^c_{\bar L} $ & $(2n+2)/R$ \\
\hline
\end{tabular}
\end{center}}
\vspace{3mm}
Table 2. Parity assignments and KK masses of fields in the 4d chiral
supermultiplets resulting from the decomposition of 5d hypermultiplets in
the $(T+\bar{F})$ representation. The subscript labels the SM transformation
properties, \eg, $Q=({\bf 3},{\bf 2})_{\bf 1/6}$,
$\bar{Q}=(\bar{{\bf 3}},\bar{{\bf 2}})_{-{\bf 1/6}}$,
$\bar{U}=(\bar{\bf 3},{\bf 1})_{-{\bf 2/3}}$, etc.

We immediately note from Table~2 that the zero mode structure
(the $n=0$ components of the $(+,+)$ fields) do not fill out
a full generation of SM matter.  We just get
${\bar U}=(\bar{{\bf 3}},{\bf 1})_{-{\bf 2/3}}$,
${\bar E}=({\bf 1},{\bf 1})_{\bf 1}$, and
$L=({\bf 1},{\bf 2})_{-{\bf 1/2}}$ left-chiral
N=1 superfields.  This is rectified by taking another copy of
${\bf 10} + {\bf \bar{5}}$ in the bulk (with N=1 chiral
components denoted $T'+\bar{F'}+T'^c + \bar{F'}^c$), and using the freedom
to flip the overall action of the $P'$ parity on these multiplets by a
sign relative to $T+\bar{F}+T^c + \bar{F}^c$.\footnote{Hall and
Nomura make this observation in the 2nd footnote of the final
version of~\cite{HN}.}
This difference leads to a different selection of zero mode components,
the KK spectrum being given in Table~3.

{\begin{center}
\begin{tabular}{|c|c|c|}
\hline
$(P,P')$ & 4d superfield & 4d mass\\
\hline
$(+,+)$ &  $T'_{Q}, \bar{F'}_{\bar D} $ & $2n/R$ \\
$(+,-)$ &  $T'_{\bar U}, T'_{\bar E}, \bar{F'}_{L}$  & $(2n+1)/R$\\
$(-,+)$ &  $T'^c_{U}, T'^c_{E}, \bar{F'}^c_{\bar L} $ & $(2n+1)/R$ \\
$(-,-)$ &  $T'^c_{\bar Q}, \bar{F'}^c_{D}$ & $(2n+2)/R$\\
\hline
\end{tabular}
\end{center}}
\vspace{3mm}
Table 3. Parity assignments and KK masses of fields in 4d chiral
supermultiplets resulting from the decomposition of 5d hypermultiplets
in the $(T'+\bar{F'})$ representation. The alteration of the action
of $P'$ causes an effective interchange of the 1st and 2nd rows,
and of the 3rd and 4th rows, relative to Table~2.

Now a marvellous thing has happened!  Combining the results of Tables
2 and 3 we have zero modes which fill out the full matter content of
a SM generation.  Remarkably, this occurs in such a way as to explain
charge quantization and the hypercharge assignments despite the fact
that what we think of as a single generation
filling out a ${\bf 10} + {\bf \bar{5}}$, has components that
arise from {\em different} $({\bf 10} + {\bf \bar{5}})$'s in the
higher-dimension. This means that when we couple three copies (for the
three generations) of this combination of fields
to the Higgs on the $O'$ brane at $y=\ell$, {\em the Yukawa couplings
do not have to satisfy the usual SU(5) relations}.\footnote{These
results are reminiscent of those from studies of string theory
carried out in the mid-to-late 1980's~\cite{string}.
We also thank Lawrence Hall for discussions concerning
Yukawa couplings.}  Explicitly, the Yukawa couplings for the
light zero-mode MSSM matter fields now result from 3 {\em different}
combinations of the 5d fields,
\beq
W_{\rm Yukawa} =
\int d^2\th \de(y-\ell) \sum_{IJ}\biggl(
h^u_{IJ} T^{'I}_Q T^J_{\bar U} H_u+
h^d_{IJ} T^{'I}_Q {\bar F'}^{J}_{\bar D} H_d+
h^l_{IJ} T^{I}_{\bar E} {\bar F}^{J}_{L} H_d \biggr),
\label{eq:YukII}
\eeq
where we have indicated the zero mode
components contained in the 5d $T, T', {\bar F'}$ and ${\bar F}$
fields (using the same notation for the components
as in Tables~2 and 3).  As before, $H_u$ and $H_d$ are 4d chiral fields
localized to the $O'$ brane at $y=\ell$.  Thus we have no
problems with the standard incorrect 1st and 2nd generation
mass-ratio predictions of SU(5), and do not require the contortions
of other models to avoid these difficulties.

Having discussed how a successful understanding of quantum numbers
and interactions may be achieved, we now turn to the question of the
nature of gauge coupling unification in these models.

\section{Gauge Unification at Tree Level and Beyond}

At least in the domain of applicability of quantum (effective)
field theory, the gauge couplings in an $S^1/(Z_2\times Z_2')$ orbifold
GUT of the general type discussed here and in Refs.~\cite{kaw1,kaw2,kaw3,AF,
HN} never {\em exactly} unify. The reason for this is that the theory on the
GUT-breaking brane $O'$ never recovers the full SU(5)
symmetry as the energy scale is increased. Its symmetry remains limited to
the SM gauge group. In particular, there are brane-localized 4d kinetic
terms for the SM gauge fields with SU(5)-violating coefficients $1/g_i^2$
(see Eq.~(\ref{eq:effL})). Nevertheless, a 4d observer performing experiments
at or just below the energy scale $M_c=1/R$ sees an approximate SU(5)
unification of couplings. The reason for this is that the bulk gauge
kinetic term is SU(5)-symmetric and the wavefunction of the zero-mode
gauge boson is independent of $y$.  Thus the physical gauge
coupling, as measured in such experiments, is obtained by
integrating over the 5th dimension, and if the linear extent of the
5th dimension is sufficiently large (we shall
quantify this shortly), then the SU(5)-violating brane kinetic terms are
dominated by the bulk contribution, leading to an approximate
SU(5)-symmetry.

This can be made more explicit by developing the discussions of Hall and
Nomura, and Nomura, Smith and Weiner~\cite{HN,NSW}.
We begin by estimating the region of validity of the
5d field-theoretic description.
Disregarding, for the moment, the $\de$-function terms in
Eq.~(\ref{eq:effL}) and integrating over the fifth
dimension, one finds that the 5d gauge coupling $g_5$ is related to the 4d
gauge couplings of the SM by $g_1^2=g_2^2=g_3^2=g_U^2=g_5^2/\ell$.
Let us first assume that this tree-level relation is approximately valid at
scale $1/R$. Since the couplings $g_i$ run in a conventional way below that
scale, this puts $1/R$ in the vicinity of $M_{GUT}\sim 10^{16}$ GeV.
Phenomenologically, $\alpha_U\simeq g_U^2/4\pi \simeq 1/25$. We now show
that the smallness of this number allows a large region of validity of
weakly coupled 5d gauge theory. By na\"{\i}ve dimensional analysis (NDA)
(cf. Ref~\cite{CLP}), the effective loop expansion parameter in
D dimensions is given by the `reduced coupling'
$\bar{g}^2=g^2/(2^D\pi^{D/2}\Gamma(D/2))$.
This means, in particular, that $\bar{g}_4^2=g_4^2/(16\pi^2)$ in 4d
and $\bar{g}_5^2=g_5^2/(24\pi^3)$ in 5d. The 5d gauge theory, being
non-renormalizable, has power-divergent loop corrections. With a cutoff
$\Lambda$, these loop corrections are negligible as long as $\bar{g}_5^2
\Lambda\ll 1$. Thus, the 5d theory becomes strongly interacting at the scale
\beq
\La\simeq \frac{1}{\bar{g}_5^2}\simeq \frac{12\pi}{\alpha_U}\,R^{-1}
\simeq 10^3\,R^{-1}\,.
\label{eq:cutoff}
\eeq
We see that the range of validity of weakly coupled 5d gauge theory can span
up to three orders of magnitude and may lead us directly to, \eg,
string physics at the Planck scale.


We now turn to loop corrections to the values of $g_i$ at the scale
$M_c\equiv1/R$. One
approach is to integrate out the fifth dimension at the maximal scale $M$
at which our model is valid and to run the 4d effective theory down to
$M_c=1/R$ in a conventional manner. (Note that $M$ can be significantly
smaller than its obvious upper bound $\La$ if, for example, the internal
structure of the brane is resolved at a lower energy.) The 4d gauge
couplings are obtained by applying $g_i^2=g_5^2/\ell$ (for $i=1,2,3$)
at scale $M$ and integrating out the KK modes between $M$ and $M_c$:
\beq
\alpha_i^{-1}(M_c)=\alpha_U^{-1}+\frac{1}{2\pi}\left\{a_i\ln\frac{M}{M_c}+
b_i\sum_{n=1}^N\ln\frac{M}{(2n-1)M_c}+c_i\sum_{n=1}^N\ln\frac{M}{2n\,M_c}
\right\}\,.\label{run}
\eeq
Here $a_i$ is the zero-mode contribution (corresponding to the
usual MSSM running) while $b_i$ and $c_i$ are the contributions
of the higher KK modes. The masses of these modes are even or odd multiples
of $M_c$, depending on the boundary conditions at $O$ and $O'$. The
upper limit of the sum is determined by the UV cutoff $N \sim M/2M_c$.
Rewriting Eq.~(\ref{run}) as
\beq
\alpha_i^{-1}(M_c)=\alpha_U^{-1}+\frac{1}{2\pi}\left\{a_i\ln\frac{M}{M_c}+
(b_i+c_i)\sum_{n=1}^N\ln\frac{M}{2nM_c}-b_i\sum_{n=1}^N\ln\left(1-
\frac{1}{2n}\right)\right\},\label{run1}
\eeq
it seems impossible to extract any useful information about low-energy
gauge couplings since the sum multiplying $(b_i+c_i)$ produces a
power-like dependence on the cutoff $M$. The reason is that a power-like
cutoff dependence has an error which is not parametrically smaller than the
effect itself. Note however that this power-like behaviour is a
manifestation of the 5d SU(5)-{\em symmetric} bulk. Thus,
even without a detailed calculation, we know the sum $(b_i+c_i)$ to be
independent of $i$. As a result, the physically important
differences $\alpha_{ij}(M_c)\equiv\alpha_i^{-1}(M_c)-\alpha_j^{-1}(M_c)$
are only logarithmically sensitive to the cutoff scale and are therefore
calculable:
\beq
\alpha_{ij}(M_c)=\frac{1}{2\pi}\left\{a_{ij}+\frac{1}{2}b_{ij}\right\}
\ln\frac{M}{M_c}\,.\label{dr}
\eeq
Here $a_{ij}=a_i-a_j$, $b_{ij}=b_i-b_j$, and the sum multiplying $b_i$ in
Eq.~(\ref{run1}) has been evaluated at leading logarithmic accuracy.
The effect described by Eq.~(\ref{dr}) has been called `differential
running' in~\cite{NSW}, where it was discussed in the framework of GUT
breaking by an adjoint Higgs on the brane.

Let us now compare this to the running of the 5d
effective action from $M$ to $M_c$. Due to unbroken gauge invariance in the
bulk, the most general form of the effective lagrangian (disregarding SM
matter) at scale $\mu$ (with $M\gg\mu\gg M_c)$ is given by
Eq.~(\ref{eq:effL}). This lagrangian, which is obtained from the original
lagrangian ${\cal L}_{\rm eff}(M)$ by integrating out all modes with
momenta above $\mu$, has to be evaluated on fields that are smooth on a
scale $\mu$. Thus, the $\delta$ functions in Eq.~(\ref{eq:effL}) have to be
understood as approximate $\delta$ functions, localized on a scale $\mu$.

An explicit calculation of ${\cal L}_{\rm eff}(\mu)$ on the basis of ${\cal
L}_{\rm eff}(M)$ is not straightforward since it requires a gauge-invariant
separation of high- and low-momentum gauge fields. Nevertheless, some
information about ${\cal L}_{\rm eff}(\mu)$ can be obtained by deriving
differences of low-energy gauge couplings on the basis of
Eq.~(\ref{eq:effL})\footnote{
A careful treatment using effective field theory with dimensional
regularization is possible~\cite{CPR}. We thank R.~Rattazzi for
discussions of this issue.
}.
For example, the analog of Eq.~(\ref{dr}), derived from Eq.~(\ref{eq:effL}),
reads
\beq
\alpha_{ij}(M_c)=\alpha_{ij}(\mu)\,+\,\frac{1}{2\pi}\left\{a_{ij}+\frac{1}
{2}b_{ij}\right\}\ln\frac{\mu}{M_c}\, .
\eeq
At the same time, one can start
from the original lagrangian ${\cal L}_M$ and derive
\beq
\alpha_{ij}(M_c)=\alpha_{ij}(M)\,+\,\frac{1}{2\pi}\left\{a_{ij}+\frac{1}
{2}b_{ij}\right\}\ln\frac{M}{M_c}\,.
\eeq
This implies
\beq
\alpha_{ij}(\mu)=\alpha_{ij}(M)\,+\,\frac{1}{2\pi}\left\{a_{ij}+\frac{1}
{2}b_{ij}\right\}\ln\frac{M}{\mu}\,,\label{brun}
\eeq
thus determining the evolution of the $SU(5)$-breaking brane-terms in
Eq.~(\ref{eq:effL}) with $\mu$.

The differences $\alpha_{ij}(M)$ in Eq.~(\ref{brun}) are parameters of
the (for our purposes most fundamental) action at scale $M$. Since
Eq.~(\ref{brun}) holds only in the leading logarithmic approximation, we
have to ascribe an uncertainty of approximately $\pm 1$ to the expression
$\ln(M/\mu)$. Equivalently, this uncertainty can be attributed to the value
of $\alpha_{ij}(M)$. This implies that it is unnatural for the quantity
$\alpha_{ij}(M)$ (which corresponds to a coefficient of a term in the
effective lagrangian at scale $M$) to be smaller than this renormalization
scale uncertainty. In the most optimistic scenario, where no new physics at
scale $M$ contributes to $\alpha_{ij}(M)$, we have therefore to assume that
\beq
\alpha_{ij}(M)\sim\frac{1}{2\pi}\left\{a_{ij}+\frac{1}{2}b_{ij}\right\}\,,
\label{err}
\eeq
where ``$\sim$'' means that an ${\cal O}(1)$ coefficient remains
undetermined.\footnote{
In Ref.~\cite{HN} the size of these brane localized gauge-kinetic terms is
estimated to be $\alpha_i^{-1}\sim 1/(4\pi)$ using NDA and the assumption
that the 4d gauge theory on the boundary becomes strongly interacting at
the same scale as the 5d bulk theory. Our result, which is similar
numerically, does not rely on this assumption and is also valid in the case
of a weakly coupled 4d theory.}

To develop some intuition for the numerical values of the relevant scales,
focus first on model I of Sect.~2.3. The relevant $\beta$ function
coefficients are $a_i=(33/5,1,-3)$ (MSSM) and $b_i=(-10,-6,-4)$,
$c_i=(0,-4,-6)$ (KK modes) (cf.~\cite{HN,NSW}). With $\alpha_i^{-1}(m_Z)=
(59.0,29.6,8.4)$ and an effective SUSY breaking scale $m_Z$, conventional
unification occurs at $M_{GUT}\simeq 2\times 10^{16}$ GeV. An obvious
possibility is to have $M_c$ only slightly below $M_{GUT}$, in which case
unification (within the uncertainty specified by Eq.~(\ref{err})) occurs
slightly above $M_{GUT}$ (since above $M_c$ the $\alpha_{ij}$ run somewhat
slower than in the MSSM). In this scenario, even though $\Lambda$ is near
the Planck scale, the scale $M$ at which couplings unify and our orbifold
model breaks down is near $M_c$. In the opposite extreme, one can have
$M_c\simeq 2\times 10^{14}$ GeV and unification at $M\simeq 2\times 10^{17}$
GeV. In this scenario, the 5d orbifold model is valid over 3 orders of
magnitude and the 5d Planck scale is reached. In model II of Sect.~2.3, the
zero-mode $a_i$ coefficients are identical to model I, while the
non-zero KK mode $\beta$ function coefficients are $b_i=(2,6,8)$, and
$c_i=(12,8,6)$.  Note that not only is the
difference $a_{ij}=a_i-a_j$ the same for the two models, but so
is $b_{ij}=b_i-b_j$, namely $(b_{12},b_{23},b_{31})=
(-4,-2,6)$.  The reason for this is that the alteration between
the two models is that model II has, effectively, at every non-zero KK level
a number of complete SU(5) multiplets (as can be seen by combining
the quantum numbers at each non-zero-mode mass level from Tables~2 and 3).
Thus the differences $b_{ij}$ are determined solely by the gauge
structure of the bulk which is unchanged.  Therefore, apart from a somewhat
reduced cutoff scale due to the faster SU(5)-invariant running of the bulk
theory, the two models are very similar in this regard.

Summarizing, the
primary phenomenological result of this section is that in both models I
and II the uncertainty in the gauge coupling unification (quantified by
Eq.~(\ref{err})) is comparable to the GUT scale threshold corrections of a
conventional GUT.

\section{Conclusions}

In this letter we have investigated supersymmetric SU(5) grand-unified
theories (GUTs)
realized in 5 space-time dimensions and broken down to the MSSM by
SU(5)-violating boundary conditions on a $S^1/(Z_2\times Z_2')$ orbifold
with two `end-of-the-world' 3-branes.  Because of the position dependence
of the amount of gauge symmetry, the MSSM Higgs doublets can be located
on the brane on which SU(5) is not a good symmetry.  Thus SU(3) triplet
partners are never required, and the doublet-triplet splitting
problem is entirely avoided.  In Sect.~2.3 we constructed
an extremely simple model in which the MSSM matter is also located on
this SU(5)-violating brane.  In this case, despite the loss of the
SU(5) understanding of the multiplet structure of the MSSM matter,
one maintains gauge coupling unification up to small corrections.
As we discuss in Sect.~3, this is due to the $y$-independence of the
SM gauge boson zero modes, which allow the contribution of the
SU(5)-symmetric bulk gauge kinetic term to dominate that of the
SU(5)-violating brane-localized kinetic terms.  We find it intriguing
that such a simple model is consistent.  In Sect.~2.3 we also
constructed a second model, this time with MSSM matter located in the
SU(5)-symmetric bulk.  The model is slightly more involved, but
preserves the SU(5) predictions for fermion quantum numbers as well
as gauge-coupling unification.  Both models
naturally avoid problematic SU(5) predictions for the Yukawa couplings of
the first two generations. Finally, in Sect.~3, we analysed the running
of the gauge couplings above the compactification scale in terms of a
5d effective action and derived the implications for the values of
the compactification scale, the unification scale and of the scale
at which the bulk gauge theory becomes strongly coupled, finding values
that are phenomenologically attractive.

\noindent
{\bf Acknowledgements}:
We are very grateful to Guido Altarelli, Wilfried Buchm\"uller,
Lawrence Hall, and Riccardo Rattazzi for helpful conversations.
JMR thanks the members of the Theory Group, LBNL, for their
kind hospitality during the completion of this work.

\end{document}